# Dissimilarity learning via Siamese network predicts brain imaging data

Aakash Agrawal

Centre for BioSystems Science & Engineering

Indian Institute of Science, Bangalore, 560012, India

*Correspondence to: Aakash Agrawal (aakash@iisc.ac.in)

## ABSTRACT

The advent of deep learning has a profound effect on visual neuroscience. It paved the way for new models to predict neural data. Although deep convolutional neural networks are explicitly trained for categorization, they learn a representation similar to a biological visual system. But categorization is not the only goal of the human visual system. Hence, the representation of a classification algorithm may not completely explain the visual processing stages. Here, I modified the traditional Siamese network loss function (Contrastive loss) to train them directly on neural dissimilarity. This network takes image pair as input and predicts the correlation distance between their output features. For Algonauts challenge, using dissimilarity learning, I fine-tuned the initial layers of Alexnet to predict MEG early response/EVC data and all the layers of VGG-16 to predict MEG late response/IT data. This approach is ideal for datasets with high SNR. Therefore, my model achieved state-of-the-art performance on MEG dataset but not fMRI.

## INTRODUCTION

Biological vision has inspired computer vision algorithms over decades. At the same time, advances in computer vision yielded better models to predict brain activity. Neurophysiological studies from primary visual cortex suggest that neuron in V1 responds to oriented bars (HUBEL & WIESEL, 1962). These inspired early models of computer vision such as Gabor filter bank, SIFT, HOG etc. which were followed by hierarchical models like Hmax (Riesenhuber & Poggio, 1999) to predict neural responses in higher visual areas. Recent deep convolutional neural networks models have achieved remarkable performance on various visual tasks. Interestingly, the features learnt by these networks match the visual representation of objects in humans. Also, there is an equivalence in the visual representation of objects between layers of deep networks and stages of biological visual processing (Cadieu et al., 2014; Guclu & van Gerven, 2015; Khaligh-Razavi & Kriegeskorte, 2014). Also, the accuracy of deep neural networks can be improved by constraining its representation to match perceptual representation (McClure & Kriegeskorte, 2016). Hence, there is a need to build models to predict the representation space (Cichy et al., 2019).

Contrary to our visual systems, CNNs trained for categorization task require millions of training samples and cannot generalize to new categories without retraining it. In an alternate approach, instead of learning a categorization model, one can learn a distance model such that images from the same category are close to each other and farther away from the images of another category. This is achieved by using a discriminative loss function on a Siamese network (Chopra, Hadsell, & LeCun, 2005). The discriminate loss function maximizes the inter-class distance but does not preserve the perceptual space in humans (Arun, 2012; Kriegeskorte, Mur, & Bandettini, 2008). Hence, I propose a new loss function to train a Siamese network directly on dissimilarity values. This training regime achieved the state-of-the-art results on predicting MEG dissimilarities values of the test dataset in the Algonauts challenge (Cichy et al., 2019). In this approach, I fine-tuned a pretrained AlexNet (Krizhevsky, Sutskever, & Hinton, 2012) and VGG-16 (Simonyan & Zisserman, 2014) models using the neural dissimilarity values of fMRI and MEG training datasets.

# METHODS

In this study, I trained a Siamese network (https://github.com/lenck/siamese-mnist) to minimize the difference between correlation distance estimated from a given layer and the observed dissimilarity value. Siamese network is made up of two parallel networks that share the same weights (Fig 1). Let X1 and X2 be the input images to the parallel networks. Since the parallel networks have common weight, the output from a given layer is f(X1) and f(X2). Here, 'f' is a function that transforms the input image into a feature vector.

The dissimilarity between an image pair is calculated as $dpred = 1 - Pearson\ correlation\ (f(X1), f(X2))$. The loss (L) between predicted and observed dissimilarity is estimated using squared Euclidean distance i.e. $L(observed, predicted) = (|| dpred - dobs ||)^2$. Note that the loss between observed and predicted dissimilarity can also be estimated using correlation distance but was not used in this study for simplicity. The loss function L was minimised via stochastic gradient descent.

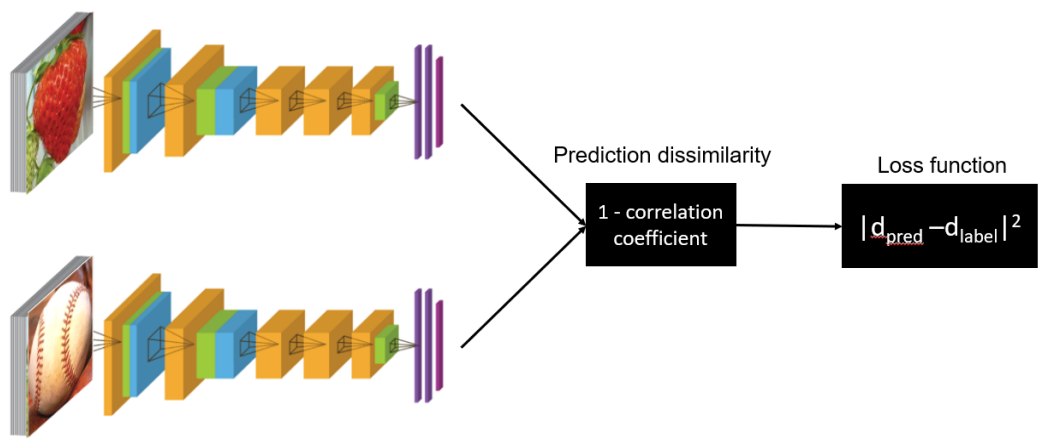

**Figure 1: Schematic of Siamese network**

Siamese architecture includes two identical networks that have common weights. Dissimilarity between the feature of an image pair can be estimated from any layer. The predicted dissimilarity is then minimized to match the given label. This figure is modified from (Cichy et al., 2019)

**Track 1 – fMRI dataset**

The training set was comprised of 2 datasets containing 92 and 118 images, respectively. Pair-wise dissimilarity values averaged across subjects were concatenated into a single training set. Prior to averaging, I normalized the dissimilarity value for each subject using MATLAB function '*zscore*'. Thus, the training data labels contained 11089 pair-wise dissimilarity values ($^{92}C_2 + ^{118}C_2$).

The early RDMs were estimated by finetuning the weights of pretrained AlexNet via Siamese network. The parameters of later (from conv5) layers were not changed during the training process. I used a fixed learning rate of 0.005 and a batch size of 32. The late RDMs were estimated by finetuning the weights of pretrained VGG-16 via Siamese network using a fixed learning rate of 0.005 and a batch size of 32. The parameters of initial (up to conv5_2) layers were not changed during the training process. To predict the test labels, I deleted the parallel network and used the correlation distance from the layer and epoch that best matched the test RDMs.

**Track 2 – MEG dataset**

The training set was comprised of 2 datasets containing 92 and 118 images, respectively. Pair-wise dissimilarity values averaged across subjects and 20-time points were concatenated into a single training set. Prior to averaging, I normalized the dissimilarity value for each subject using MATLAB function '*zscore*' and transformed it back to approximately the same range of dissimilarities i.e. x = (x+2)/5. Thus, the training data labels contained 11089 pair-wise dissimilarity values ($^{92}C_2$ + $^{118}C_2$).

The early RDMs were estimated by finetuning the weights of pretrained AlexNet via Siamese network using a fixed learning rate of 0.005 and a batch size of 16. The parameters of later layers (from conv5) were not changed during the training process. The late RDMs were estimated by finetuning the weights from all layers of pretrained VGG-16 via Siamese network using a fixed learning rate of 0.005 and a batch size of 16. To predict the test labels, I deleted the parallel network and used the correlation distance from the layer and epoch that best matched the test RDMs. The input image of the test dataset for MEG late was resized to 300x300.

## RESULTS AND DISCUSSION

It is known that the representation of early visual cortex (EVC) is dependent on low-level stimulus properties, and the representation of higher visual areas (IT) is more categorical. While the receptive field of neurons in early visual areas are small, each voxel in fMRI contains millions of such neurons. Hence, it has a large receptive field and encodes information from a large part of an image (Kay, Naselaris, Prenger, & Gallant, 2008). Therefore, I finetuned the weights of Alexnet to train EVC/ MEG-early response as it has the largest filter size (in the first layer) among the existing architectures. But Alexnet has relatively low classification accuracy and also did not yield good model predictions. Therefore. I finetuned the weights of VGG-16 to predict the dissimilarities of IT/MEG-late response. This approach explained 63.56% of explainable variance in MEG dataset and 24.03% of explainable variance in fMRI dataset. To predict MEG Representation Dissimilarity Matrix (RDM), I used layer 12 from epoch 164 of finetuned Alexnet to predict MEG_early_RDMs and layer 34 from epoch 13 of finetuned VGG-16 to predict MEG_late_RDMs. Similarly, in fMRI dataset, I used layer 10 from epoch 250 to predict EVC_RDMs and layer 34 from ep 116 to predict IT_RDMs.

Since the number of image pairs is orders of magnitude higher than the number of unique images, I was able to fine-tune the existing convolution networks without any external dataset. But performance on fMRI dataset is low because of low SNR and might improve by using RDMs from external datasets. Further, it will be interesting to know if the accuracy of networks trained to match perception also leads to better classification performance but that is beyond the scope of this paper.

## ACKNOWLEDGEMENT

I would like to thank the IISc VisionLab computational resources that helped me train my models for this competition.